\documentclass[aps,pra,twocolumn,groupedaddress,nofootinbib]{revtex4-2}
\usepackage{amsmath}
\usepackage{amsthm}
\usepackage{amssymb}
\usepackage{graphicx}
\usepackage{mathrsfs}
\usepackage{mathtools}
\usepackage{stmaryrd}
\usepackage{enumerate}
\usepackage{enumitem}
\usepackage[T1]{fontenc}
\usepackage{physics}
\usepackage{subfigure}
\usepackage{overpic}
\usepackage{epstopdf}
\usepackage{bm}
\usepackage{color,soul}
\usepackage{tikz}
\usepackage[bookmarks=true]{hyperref}
\usepackage{xcolor}
\usepackage[linesnumbered]{algorithm2e}

\hypersetup{colorlinks=true,citecolor=blue,
linkcolor=blue,urlcolor=blue,pdfstartview=FitH,
bookmarksopen=true}

\allowdisplaybreaks

\usepackage{cleveref}
\crefname{equation}{Eq.}{Eqs.}
\Crefname{equation}{Equation}{Equations}
\crefname{theorem}{Theorem}{Theorems}
\crefname{lemma}{Lemma}{Lemmas}
\crefname{appendix}{Appendix}{Appendixes}
\crefname{figure}{Fig.}{Figs.}
\crefname{section}{Sec.}{Secs.}
\Crefname{section}{Section}{Sections}
\crefname{algorithm}{Algorithm}{Algorithms}

\begin{document}

\title{Unified and computable approach to optimal strategies for multiparameter estimation}
\author{Zhao-Yi Zhou}
\affiliation{Department of Physics, Shandong University, Jinan 250100, China}
\author{Da-Jian Zhang}
\email{zdj@sdu.edu.cn}
\affiliation{Department of Physics, Shandong University, Jinan 250100, China}
\date{\today}

\begin{abstract}
Precise estimation of physical parameters underpins both scientific discovery and technological development. A central goal of quantum metrology and sensing is to exploit quantum resources like entanglement to devise optimal strategies for estimating physical parameters as precisely as possible. While substantial progress has been made in single-parameter quantum metrology, the multiparameter scenario remains significantly more challenging due to the issue of parameter incompatibility. In this work, we present a unified and computable approach for the simultaneous estimation of multiple parameters that attains the ultimate precision permitted by quantum mechanics. The core of our approach is to integrate the quantum tester formalism into the recently proposed tight Cram\'er-Rao type bound. This formulation enables us to figure out the highest achievable precision via upper and lower bounds that are computable via semidefinite programs. More importantly, within this formulation, diverse quantum resources, including entanglement, coherence, quantum control, and indefinite causal order, are treated on equal footing and systematically optimized for the purpose of achieving the ultimate precision in multiparameter estimation. As a result, our approach is applicable to various metrological strategies both in the presence and absence of noise. To demonstrate its utility, we revisit three-dimensional magnetic-field estimation, uncovering the strengths and limitations of existing analytical results and further establishing a strict hierarchy among different types of strategies.
\end{abstract}

\maketitle
\section{Introduction}
Much of quantitative science deals with measuring unknown parameters of a physical process. With the advent of quantum technologies, it has been found that quantum resources such as entanglement and coherence allow for pushing the measurement precision limit beyond what is achievable with classical means \cite{Giovannetti2004,Giovannetti2006}. A central goal in quantum metrology and sensing is therefore to devise optimal strategies for exploiting quantum resources to estimate parameters as precisely as possible, which has motivated vibrant research activities over the past two decades \cite{Giovannetti2011,Pezze2018RoMP}. 

The problem of devising optimal strategies has been well addressed in the single-parameter setting \cite{LZC21,LHYY24}. This is largely because the ultimate precision achievable is fully characterized by the quantum Cram\'er--Rao
bound, which can be saturated by suitable measurements
under broad conditions \cite{Braunstein1994,Zhang2015PRL,Lu2015NC,Zhang2020,ZWZ20,Xu2020PRL,Li2021PRA,Zhang2022a,Zhang2024,Zhou2025PRA,Zhang2025}. This theoretical achievement has led to a variety of experimentally relevant
protocols realizing quantum-enhanced measurements in systems ranging
from optical interferometry to atomic clocks and solid-state sensors \cite{Jiao2023,PFD25,Montenegro2025PR}.
The situation, however, changes dramatically in the multiparameter regime. When several parameters are estimated simultaneously, the optimal
measurements associated with different parameters may be mutually
incompatible, preventing the simultaneous saturation of the
single-parameter quantum Cram\'er--Rao bounds.
This issue, known as parameter incompatibility, renders the characterization of ultimate precision
substantially more subtle and has been recognized as a central obstacle
in multiparameter quantum metrology \cite{Lu2021PRL,Albarelli2022PRX}.
As such, identifying both the fundamental precision limits and
the corresponding optimal strategies remains an open challenge in multiparameter scenarios.

The past few years have witnessed considerable efforts devoted to addressing this challenge, driven by the growing relevance of multiparameter estimation in applications like quantum imaging \cite{TNL16,LP16,CDJB17,CDJB17,RHS17}, magnetic field sensing \cite{BD16,HZX20,ZCHL24}, and Hamiltonian parameter estimation \cite{Yua16}. It has become increasingly clear that quantum-enhanced measurements may be enabled by a variety of quantum resources, such as entanglement \cite{Giovannetti2004}, coherence \cite{Giovannetti2006}, quantum control \cite{Yua16}, and even indefinite causal order \cite{ZYC20,LHYY23}. While each of these ingredients has been demonstrated to provide advantages in specific scenarios, their roles in multiparameter estimation are often analyzed in isolation, making a systematic and unified treatment elusive. Furthermore, the presence of noise, which is ubiquitous in practical applications, substantially complicates the identification of truly optimal strategies for multiparameter estimation. Therefore, a unified framework capable of systematically optimizing over
all available quantum resources is still lacking in multiparameter scenarios up to now.

In this work, we present a unified and computable approach for the simultaneous estimation of multiple parameters that attains the ultimate precision permitted by quantum mechanics. Our approach is grounded in the tight Cram\'er-Rao type bound (TCRB) recently introduced by Hayashi and Ouyang \cite{HO23}, which fully characterizes the highest achievable precision in estimating multiple parameters from a given output state. The new contribution of our work lies in integrating the quantum tester formalism \cite{CDP08,CDP09,ABC15} into the TCRB. Specifically, we formulate the problem of identifying the highest achievable precision as a conic programming problem, where the optimization is performed over all admissible quantum testers within a specified type of strategies. This formulation enables us to figure out the highest achievable precision via upper and lower bounds that are computable via semidefinite programs (SDPs). More importantly, within this formulation, diverse
quantum resources, including entanglement, coherence, quantum control, and indefinite
causal order, are treated on equal footing and systematically optimized for the purpose of achieving the ultimate precision in multiparameter estimation. As a result, our approach is applicable to various metrological
strategies both in the presence and absence of noise. 

We clarify that, although the derivations presented below closely follow those in Ref.~\cite{HO23}, the results obtained here are fundamentally interesting and bear significant physical implications. Indeed, the TCRB was formulated to find optimal measurements given a fixed output state, without considering how this state is generated \cite{HO23,Note}. Consequently, the TCRB alone cannot determine the optimal strategies, as it does not account for how the output states are generated. In contrast, by incorporating quantum testers into the TCRB, our approach provides a complete characterization of the entire estimation protocol, encompassing state preparation, parameter encoding, intermediate operations, and final measurements. As such, all quantum resources are fully taken into account in our approach. To demonstrate the
physical relevance of our approach, we apply it to the three-dimensional
magnetic-field estimation \cite{BD16,HZX20}, a paradigmatic and fundamentally
important task in quantum metrology. We first analyze the noiseless setting and show that existing analytical results exhibit both notable strengths and intrinsic limitations. We then turn to the noisy regime and establish a strict
hierarchy among different types of strategies. Our approach thus represents
a valuable tool capable of providing deeper physical insight into multiparameter estimation.

This paper is organized as follows. In \cref{sec_preliminaries}, we recall the quantum tester formalism. In \cref{sec_nearoptimal}, we present the unified formulation with the aid of quantum testers. In \cref{sec_upperbound} and \cref{sec_lowerbound}, we derive upper and lower bounds, respectively. We then revisit analytical results for the three-dimensional magnetic field estimation in \cref{sec_application_1} and establish a strict hierarchy among different types of estimation strategies in \cref{sec_application_2}. We conclude this work in \cref{sec_conclude}.

\section{Preliminaries} \label{sec_preliminaries}
\begin{figure}[]
    \centering
    \includegraphics[width=\linewidth]{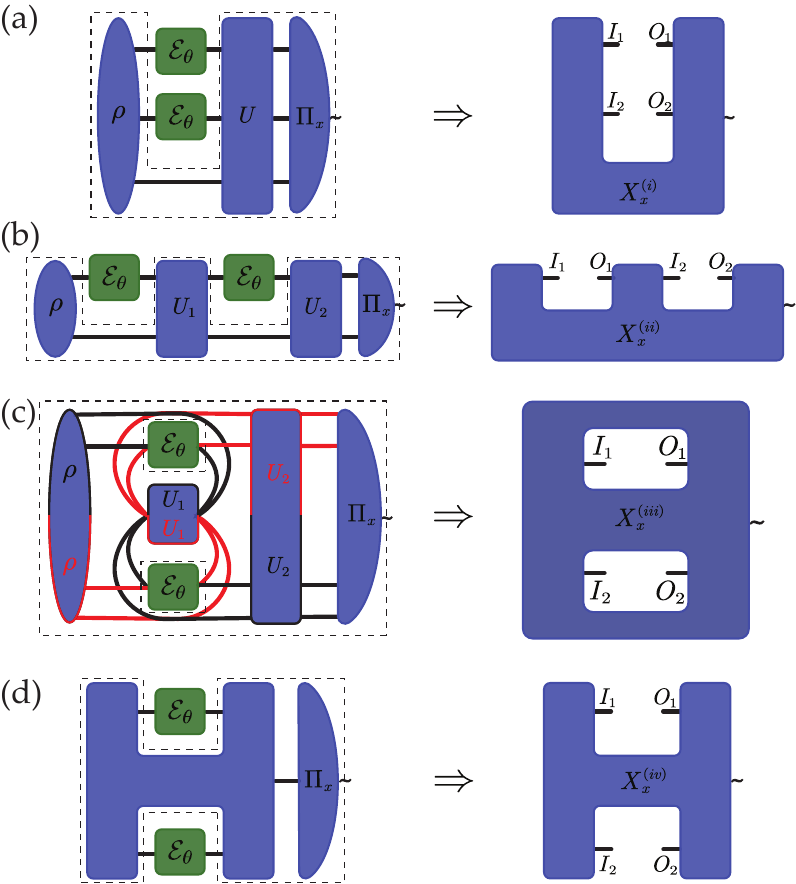}
    \caption{Schematic of quantum testers with $N=2$. The dashed boxes in the left panels are concrete realizations of estimation strategies: (a) parallel strategies, (b) sequential strategies, (c) causal superposition strategies, and (d) general indefinite-causal-order strategies. Here $\rho$ is the input state, $\mathcal{E}_{\bm{\theta}}$ represents the parameter-encoding channel, and $U_i$ denotes the unitary control operations. In each of these strategies, a final measurement $\{\Pi_x\}_x$ is performed on the output state. The black and red curves in (c) represent two different causal orders. The right panels display the associated quantum testers.}
    \label{fig_testers}
\end{figure}

To present our results clearly, we need to recapitulate the quantum tester formalism \cite{CDP08,CDP09,ABC15}. Suppose that there are $N$ identical quantum channels $\mathcal{E}_{\bm{\theta}}$ at our disposal. Here, $\boldsymbol{\theta}=(\theta_1,\theta_2,\cdots,\theta_\mathfrak p)^T$ collectively denotes the unknown parameters, where $\mathfrak{p}$ is the number of parameters in question.
Let $\mathcal{H}_{I_k}$ and $\mathcal{H}_{O_k}$ represent the input and output Hilbert spaces of the $k$th copy of the channel, respectively. 
Denote by $\mathcal L(\mathcal H)$ the set of linear operators over a Hilbert space $\mathcal H$. The $k$th copy of the channel $\mathcal{E}_{\bm{\theta}}$, as a linear map from $\mathcal{L}(\mathcal{H}_{I_k})$ to $\mathcal{L}(\mathcal{H}_{O_k})$, can be associated with a positive semidefinite operator in $\mathcal{L}(\mathcal H_{I_k}\otimes \mathcal{H}_{O_k})$
\begin{equation}
  E_{\bm{\theta}}\coloneqq \mathrm{id}\otimes\mathcal{E}_{\bm{\theta}}\left(| I \rangle\hspace{-2.7pt}\rangle\langle\hspace{-2.7pt}\langle I|\right),
\end{equation}
known as the Choi-Jamio\l{}kowski (CJ) operator \cite{Jam72,Cho75}. Here, $\mathrm{id}$ denotes the identity map and $| I \rangle\hspace{-2.7pt}\rangle=\sum_j|j\rangle|j\rangle$. The CJ operator corresponding to the $N$ identical channels is 
\begin{equation} \label{eq_CJ_Nchannel}
C_{\bm{\theta}} \coloneqq E_{\bm{\theta}}^{\otimes N}\in \mathcal{L} \left( \mathcal{H}_{IO} \right),
\end{equation}
where $\mathcal{H}_{IO}\coloneqq \mathcal{H} _{I_1}\otimes \mathcal{H} _{O_1}\otimes \cdots \otimes \mathcal{H} _{I_N}\otimes \mathcal{H} _{O_N}$. A quantum tester is a generalization of a quantum measurement and may account for various ingredients of an estimation strategy, including the input state, quantum controls, and final measurement (see Fig.~\ref{fig_testers}). Mathematically, a quantum tester is described by
a collection of positive semidefinite operators $\{X_x\}_x$ in $\mathcal{L}(\mathcal{H}_{IO})$ such that 
\begin{equation} \label{eq_prob_tester}
    p_{\bm{\theta}}\left( x \right) \coloneqq \mathrm{tr}\left( C_{\bm{\theta}}X_{x}^T \right)
\end{equation}
are probabilities, that is, $\sum_x{p_{\bm{\theta}}(x)}=1$ and $p_{\bm{\theta}}(x)\ge 0$ for all $x$. Here, the superscript $T$ denotes the matrix transpose. 

\section{Unified formulation with quantum testers} \label{sec_nearoptimal}

To seek for the unified formulation, we exploit the quantum tester formalism to describe various types of strategies, including: ($i$) parallel strategies \cite{Giovannetti2006}, where these $N$ channels are applied simultaneously on a multipartite entangled state [see Fig.~\ref{fig_testers}(a)]; ($ii$) sequential strategies \cite{Giovannetti2006}, involving successive queries of the channels, possibly interspersed with unitary control operations [see Fig.~\ref{fig_testers}(b)]; ($iii$) causal superposition strategies \cite{ZYC20}, where the channels are probed in a superposition of different causal orders [see Fig.~\ref{fig_testers}(c)]; ($iv$) general indefinite-causal-order strategies \cite{LHYY23}, which encompass the most general causal relations among the channels and include causal superposition strategies as special cases [see Fig.~\ref{fig_testers}(d)]. 
Each type of estimation strategies corresponds to a specific set of quantum testers. Hereafter, we use index $k$ to specify the type of strategies in question, that is, $k=i, ii, iii, iv$ refer to the mentioned four types of strategies ($i$)-($iv$), respectively. We use $X_x^{(k)}$ to denote the quantum tester associated with the strategies of type $k$ and define
\begin{equation}
    X^{\left( k \right)}\coloneqq \sum_x{X_{x}^{\left( k \right)}}.
\end{equation}
It is known \cite{ABC15,Zhou2024a} that, for strategies of type $k=i,ii,iv$, the associated set of quantum testers can be characterized as  
\begin{equation} \label{eq_defBBX}
    \mathcal{X} ^{\left( k \right)}\!\coloneqq\! \left\{ X^{\left( k \right)}|X^{(k)}\!\ge\!0,\Lambda ^{\left( k \right)}\left( X^{\left( k \right)} \right) \!=\!X^{\left( k \right)},\mathrm{tr}\left( X^{\left( k \right)} \right)\!=\!d_O \right\}.
\end{equation}
Here, $d_O=\mathrm{dim}\left( \mathcal{H} _{O_1}\otimes \cdots \otimes \mathcal{H} _{O_N} \right)$ and $\Lambda ^{(k)}$ denotes a linear map from $\mathcal{L}(\mathcal{H}_{IO})$ to itself, whose explicit expression can be found in Appendix~\ref{appsec_lambda_k}. The case $k=iii$ can be treated in a similar way but with some additional care (see Appendix~\ref{appsec_striii}). 

To estimate $\bm{\theta}$, we adopt a locally unbiased estimator \cite{Helstrom1976,Holevo2011}, denoted by $\hat{\boldsymbol{\theta}}(x)$, which satisfies the two conditions 
\begin{equation} \label{eq_unbiased}
\sum_x{p_{\boldsymbol{\theta}}(x)\hat{\boldsymbol{\theta}}(x)}=\boldsymbol{\theta },
\end{equation}
\begin{equation}
\sum_x{\partial_j p_{\boldsymbol{\theta}}(x)\hat{\theta}_{i}(x)}=\delta_{ij},
\end{equation}
for $i,j=1,\cdots,\mathfrak p$. Here, $\partial_j$ stands for the partial derivative with respect to $\theta_j$, $\hat{\theta}_i(x)$ denotes the $i$-th component of $\hat{\boldsymbol{\theta}}(x)$, and $\delta_{ij}$ is the Kronecker delta. The performance of the estimator is characterized by its covariance matrix 
\begin{equation} \label{eq_cov_def}
\Sigma =\sum_x{p_{\boldsymbol{\theta}}\left( x \right) \left[ \hat{\boldsymbol{\theta}}(x)-\boldsymbol{\theta } \right] \left[ \hat{\boldsymbol{\theta}}(x)-\boldsymbol{\theta } \right] ^T}.
\end{equation}
To quantify the estimation error, we need to find a scalar function of the covariance matrix. A widely used choice is the weighted trace of the covariance matrix, $\tr(W\Sigma)$, where $W$ is a real positive semidefinite matrix that reflects the relative importance of different parameters \cite{Helstrom1976,Holevo2011}. So, we need to solve the following optimization problem:
\begin{subequations} \label{eq_opt1}
    \begin{align}
        \min\quad&\mathrm{tr}\left( W\Sigma \right) 
\\
\mathrm{s}.\mathrm{t}.\quad &\sum_x{p_{\boldsymbol{\theta}}(x)\hat{\boldsymbol{\theta}}(x)}=\boldsymbol{\theta },
\\
&\sum_x{\partial _jp_{\boldsymbol{\theta}}(x )\hat{\theta}_{i}(x)}=\delta _{ij},\quad i,j=1,\cdots ,\mathfrak{p}.
    \end{align}
\end{subequations}
Note that $p_{\boldsymbol{\theta}}(x)$ is determined by the quantum tester $\{X_x^{(k)}\}_x$ via \cref{eq_prob_tester}. The optimization in \cref{eq_opt1} is thus over all admissible quantum testers of type $k$ and all locally unbiased estimators. Note also that we can remove the first type of constraints in \cref{eq_opt1} without affecting the final result \cite{DGG20}. That is, \cref{eq_opt1} can be simplified as
\begin{subequations} \label{eq_opt2}
    \begin{align}
        \min\quad&\mathrm{tr}\left( W\Sigma \right) 
\\
\mathrm{s}.\mathrm{t}.\quad &\sum_x{\partial _jp_{\boldsymbol{\theta}}(x )\hat{\theta}_{i}(x)}=\delta _{ij},\quad i,j=1,\cdots ,\mathfrak{p}.
    \end{align}
\end{subequations}
The optimal value of \cref{eq_opt2} characterizes the highest achievable precision in estimating $\boldsymbol{\theta}$ via the strategies of type $k$.

To rewrite \cref{eq_opt2} in a convenient form, we introduce an argumented weight matrix $\tilde W$ defined over the space $\mathbb{C}^{\mathfrak{p}+1}$ 
\begin{equation}\label{eq:W-tilde}
    \tilde W \coloneqq 0\oplus W,
\end{equation}
i.e., the direct sum of $0$ and $W$. We define the vectors in $\mathbb{C}^{\mathfrak{p}+1}$
\begin{equation}\label{eq:hx}
    |h_x\rangle \coloneqq \left( \begin{array}{c}
	1\\
	\hat{\boldsymbol{\theta}}(x)-\boldsymbol{\theta }\\
\end{array} \right) .
\end{equation}
Using Eqs.~(\ref{eq:W-tilde}) and (\ref{eq:hx}) and noting that $p_{\boldsymbol{\theta}}(x)=\mathrm{tr}\left( C_{\boldsymbol{\theta }}X_{x}^{\left( k \right) T} \right)$, we can rewrite the objective function in \cref{eq_opt2} as
\begin{equation}
    \mathrm{tr}(W\Sigma) = \mathrm{tr}\left[ \tilde{W}\otimes C_{\boldsymbol{\theta }}\sum_x{|h_x\rangle \langle h_x|\otimes X_x^{(k)T}} \right] .
\end{equation}
Further, letting $A_i=\left( |0\rangle \langle i|+|i\rangle \langle 0| \right) \big /2$ with $\left\{ |i\rangle \right\} _{i=0}^{\mathfrak{p}}$ denoting the standard orthonormal basis of $\mathbb{C}^{\mathfrak{p}+1}$, we can rewrite the left-hand side of the constraints in \cref{eq_opt2} as 
\begin{equation}
    \sum_x{\partial _jp_{\boldsymbol{\theta}}(x )\hat{\theta}_{i}(x)}=\mathrm{tr}\left[ A_i \otimes \partial _jC_{\boldsymbol{\theta }}\sum_x{|h_x\rangle \langle h_x|\otimes X_{x}^{\left( k \right) T}} \right].
\end{equation}
Therefore, \cref{eq_opt2} can be equivalently formulated as 
\begin{subequations} \label{eq_opt_process}
    \begin{align}
           \underset{X_{x}^{\left( k \right)} ,\ket{h_x}}{\min}\,\,&\mathrm{tr}\left[ \tilde{W}\otimes C_{\boldsymbol{\theta }}\sum_x{|h_x\rangle \langle h_x|\otimes X_{x}^{\left( k \right)}} \right] 
           \\
\mathrm{s}.\mathrm{t}.\,\,& \sum_x{X_{x}^{\left( k \right)}}\in \mathcal{X} ^{\left( k \right)} \label{eq_res_inbbX},
\\ & X_{x}^{\left( k \right)}\ge 0,\quad \forall x, 
\\
&\mathrm{tr}\left[ A_i \otimes \partial _jC_{\boldsymbol{\theta }}\sum_x{|h_x\rangle \langle h_x|\otimes X_{x}^{\left( k \right)}} \right] =\delta _{ij},\nonumber
\\ &\quad\quad\quad\quad\quad\quad\quad\quad\quad\quad\quad i,j = 1,\cdots, \mathfrak p,
    \end{align}
\end{subequations}
where $X_x^{(k)T}$ is replaced by $X_x^{(k)}$ as $\mathcal{X}^{(k)}$ is closed under transposition. 

We observe that \cref{eq_opt_process} involves products of variables $\ket{h_x}\bra{h_x}$ and $X_x^{(k)}$, which render the optimization challenging. To address this issue, we introduce the separable cone 
\begin{equation} \label{eq_def_Ssep}
    \mathcal{S} \coloneqq \mathrm{conv}\left\{ P_1\otimes P_2|P_1\in \mathrm{Pos}\left( \mathbb{C}^{\mathfrak{p}+1} \right) ,P_2\in \mathrm{Pos}\left( \mathcal{H} _{IO} \right) \right\},
\end{equation}
where $\mathrm{Pos(\mathcal{H})}$ denotes the set of positive semidefinite operators on $\mathcal{H}$, and conv represents the convex hull \cite{BV04}. Identifying $\sum_x{|h_x\rangle \langle h_x|\otimes X_{x}^{\left( k \right)}}$ as an element $Y^{(k)}\in\mathcal{S}$, we can show that \cref{eq_opt_process} is equivalent to the optimization
\begin{subequations} \label{eq_opt_sep}
    \begin{align}
           \underset{Y^{(k)}\in \mathcal{S}}{\min}\,\,&\mathrm{tr}\left[ \tilde{W}\otimes C_{\boldsymbol{\theta }}Y^{(k)} \right]
\\
\mathrm{s}.\mathrm{t}.\,\,&\mathrm{tr}\left[ A_i \otimes \partial _jC_{\boldsymbol{\theta }}Y^{(k)} \right] =\delta _{ij}, \,\, i,j = 1,\cdots, \mathfrak p,
\\ & \mathrm{tr}_{1 }\left[ |0\rangle \langle 0|\otimes \mathbb{I} _{IO}Y^{(k)} \right] \in \mathcal{X} ^{\left( k \right)},
    \end{align}
\end{subequations}
where $\tr_1$ is the partial trace over the first subsystem and $\mathbb{I}_{IO}$ is the identity operator on $\mathcal{H}_{IO}$.
The equivalence between \cref{eq_opt_process} and \cref{eq_opt_sep} can be established as follows. Note that any feasible variable $\sum_x{|h_x\rangle \langle h_x|\otimes X_{x}^{\left( k \right)}}$ in \cref{eq_opt_process} is also a feasible variable in \cref{eq_opt_sep}. This implies that the optimal value of \cref{eq_opt_sep} is no larger than that of \cref{eq_opt_process}. On the other hand, given any feasible variable $Y^{(k)}$ in \cref{eq_opt_sep}, we can decompose it in the form
\begin{equation}\label{eq:Y}
    Y^{\left( k \right)}=\sum_x{\ket{\phi_{x}}\bra{\phi_{x}}\otimes P_{x}},
\end{equation}
where $\ket{\phi_{x}}$ is a vector in $\mathbb{C}^{\mathfrak{p}+1}$ and $P_{x}\in \mathrm{Pos}(\mathcal{H}_{IO})$. Since $A_i,|0\rangle\langle 0|$, and $\tilde{W}$ are all real symmetric matrices, only the real part of $\ket{\phi_{x}}\bra{\phi_{x}}$ contributes to the constraints and the objective function. Therefore, without loss of generality, we can assume that $\ket{\phi_{x}}$ is real. Besides, note that the components $\ket{\phi_{x}}\bra{\phi_{x}}\otimes P_{x}$ in the sum decomposition (\ref{eq:Y}) do not contribute to the constraints and the objective function in Eq.~(\ref{eq_opt_sep}) when $\langle 0|\phi_{x}\rangle = 0$. We can further assume $\langle 0|\phi_{x}\rangle \neq 0$ without loss of generality. Actually, we can assume that $\langle 0|\phi_{x}\rangle = 1$ without loss of generality, since $\ket{\phi_{x}}$ can be rescaled arbitrarily by absorbing the scaling factor into $P_{x}$. Now we can identify $\ket{\phi_{x}}$ with $|h_x\rangle$ defined in \cref{eq:hx} and $P_{x}$ with $X_{x}^{\left( k \right)}$; that is, a feasible variable $Y^{(k)}$ in \cref{eq_opt_sep} corresponds to a feasible variable $\sum_x{|h_x\rangle \langle h_x|\otimes X_{x}^{\left( k \right)}}$ in \cref{eq_opt_process}. It follows that the optimal value of \cref{eq_opt_process} is no larger than that of \cref{eq_opt_sep}. This completes the proof of the equivalence.

We highlight that Eq.~(\ref{eq_opt_sep}) is the unified formulation we seek for. Notably, once \cref{eq_opt_sep} is solved, we can determine the highest achievable precision in the strategies of type $k$, and moreover, we may construct an optimal strategy that attains this precision. Specifically, associated with an optimal solution of the form (\ref{eq:Y}), the optimal estimator is
$\hat{\theta}_{i}(x)=\theta _i+\langle i|\phi_{x}\rangle$ for $ i=1,\cdots ,\mathfrak{p} $.
Meanwhile, the optimal quantum tester can be identified as $\{ {P}_{x} \} _x$. It is worth noting that the physical realizations of quantum testers have been extensively studied. In particular, a general scheme for implementing
quantum testers corresponding to parallel, sequential, and causal
superposition strategies was developed in Refs.~\cite{CDP09,BMQ21}. Unfortunately, a concrete implementation for general indefinite-causal-order strategies remains an open problem
\cite{KGAD23,BMQ21}. Below, we show how to effectively solve \cref{eq_opt_sep} by constructing some upper and lower bounds that are computable via SDPs.

\section{Semidefinite programs for upper bounds} \label{sec_upperbound}

To construct upper bounds for \cref{eq_opt_sep}, we randomly generate $m$ unit vectors $\left\{ |w_x\rangle \right\} _{x=1}^{m}$ in $\mathbb{C}^{\mathfrak{p}+1}$. Actually, these vectors can be chosen to be real. Then, letting $Y^{(k)}=\sum_x{|w_x\rangle \langle w_x|\otimes X_{x}^{\left( k \right)}}$, where $X_{x}^{\left( k \right)}$ is a positive semidefinite matrix, we can convert \cref{eq_opt_sep} into the form
\begin{subequations} \label{eq_opt_upper}
    \begin{align}
        \underset{ X_{x}^{\left( k \right)}}{\min}\,\,&\sum_x{\langle w_x|\tilde{W}|w_x\rangle \mathrm{tr}\left( C_{\boldsymbol{\theta }}X_{x}^{\left( k \right)} \right)}
\\
\mathrm{s}.\mathrm{t}.\,\,&\sum_x{\left| \langle w_x|0\rangle \right|^2X_{x}^{\left( k \right)}}\in \mathcal{X} ^{\left( k \right)} \label{eq_res_wbbX},
\\
&X_{x}^{\left( k \right)}\ge 0,\quad \forall x,
\\
&\sum_x{\langle w_x|A_i |w_x\rangle \mathrm{tr}\left( \partial _jC_{\boldsymbol{\theta }}X_{x}^{\left( k \right)} \right)}=\delta _{ij},\nonumber
\\
&\quad \quad \quad \quad \quad \quad \quad \quad \quad \quad \quad \quad i,j=1,\cdots ,\mathfrak{p},
    \end{align}
\end{subequations} 
where the optimization variables are $\left\{ X_{x}^{\left( k \right)} \right\} _{x=1}^{m}$. Apparently, \cref{eq_opt_upper} provides an upper bound for \cref{eq_opt_sep} and different sets of $\left\{ |w_x\rangle \right\} _{x=1}^{m}$ correspond to different upper bounds. It follows from \cref{eq_defBBX} that Eq.~(\ref{eq_res_wbbX}) can be explicitly rewritten as the following two linear constraints:
\begin{equation}
    \mathrm{tr}\left[ \sum_x{\left| \langle w_x|0\rangle \right|^2X_{x}^{\left( k \right)}} \right] =d_O,
\end{equation} 
and 
\begin{equation}
    \left( \mathrm{id}-\Lambda^{(k)} \right) \left[ \sum_x{\left| \langle w_x|0\rangle \right|^2X_{x}^{\left( k \right)}} \right] =0.
\end{equation}
Therefore, the optimization problem in \cref{eq_opt_upper} is a SDP and can be solved efficiently using numerical methods \cite{GB08,CRes12}. Intuitively speaking, in the course of randomly generating more vectors $\left\{ |w_x\rangle \right\} _{x=1}^{m}$, the set of operators $\left\{\sum_x{|w_x\rangle \langle w_x|\otimes X_{x}^{\left( k \right)}}|X_{x}^{\left( k \right)}\geq 0\right\}$ becomes larger, and consequently, this set could approximate the separable cone $\mathcal{S}$ increasingly well. It is therefore expected that the optimal value of \cref{eq_opt_upper} converges to that of \cref{eq_opt_sep} as $m$ increases with a high probability. We would like to mention that similar ideas have been employed  to approximate a desired set of operators via random sampling \cite{Zhang2018}.

\section{Semidefinite programs for lower bounds} \label{sec_lowerbound}

To construct lower bounds for \cref{eq_opt_sep}, we resort to the technique of symmetric extension in entanglement theory \cite{DPS02, DPS04, TPBA24}. The main idea of this technique is to relax the set of separable states to a larger set that can be characterized by semidefinite constraints. Recall that, if $\rho$ is a separable state acting on the Hilbert space $\mathcal{H}_{1}\otimes \mathcal H_{2}$, then $\rho$ has an extension $\tilde \rho$ that acts on the extended Hilbert space $\mathcal{H}^{\otimes n}_{1}\otimes \mathcal{H}_{2}$ with the following two properties. First, $\tilde \rho$ is symmetric under interchanges of any two copies of subsystem $\mathcal H_1$. Second, tracing out any $n-1$ copies of subsystem $\mathcal H_1$ yields the original state $\rho$. These two properties can be easily verified by noting that $\tilde\rho=\sum_ip_i\rho_i^{\otimes n}\otimes \sigma_i$ if $\rho=\sum_ip_i\rho_i\otimes \sigma_i$, where $\{p_i\}$ is a probability distribution, and $\rho_i$ and $\sigma_i$ are two quantum states on $\mathcal H_1$ and $\mathcal H_2$, respectively. Moreover, if $\rho$ is entangled, there always exists one positive integer $n$ such that the above extension $\tilde \rho$ cannot be constructed. As the separate cone $\mathcal{S}$ shares the same structure as the set of separable states, we can employ the symmetric extension technique to obtain a series of converging lower bounds for \cref{eq_opt_sep}. Specifically, let $Y^{(k)}$ be a separable operator acting on the Hilbert space $\mathbb{C}^{\mathfrak{p}+1}\otimes \mathcal{H}_{IO}$. According to the symmetric extension technique, $Y^{(k)}$ has an extension $Y_n^{(k)}$ that acts on the extended Hilbert space $(\mathbb{C}^{\mathfrak{p}+1})^{\otimes n}\otimes \mathcal{H}_{IO}$ so that \cref{eq_opt_sep} can be relaxed as follows:
\begin{subequations} \label{eq_optrelax1}
    \begin{align}
        \underset{Y^{(k)}_n}{\min}\quad &\mathrm{tr}\left[ \tilde{W}\otimes C_{\boldsymbol{\theta }}Y^{(k)} \right] 
\\
\mathrm{s}.\mathrm{t}.\quad
& \mathrm{tr}_{1}\left[ |0\rangle \langle 0|\otimes \mathbb{I} _{IO}Y^{(k)} \right] \in \mathcal{X} ^{\left( k \right)},
\\
& \mathrm{tr}\left[ A_i \otimes \partial _jC_{\boldsymbol{\theta }}Y^{(k)} \right] =\delta _{ij},\,\,i,j=1,\cdots ,\mathfrak{p}, \label{eq_refbegin}
\\ 
& \big( U_{\pi}\otimes \mathbb{I} _{IO} \big) Y_{n}^{\left( k \right)}\big( U_{\pi}^{\dagger}\otimes \mathbb{I} _{IO} \big) =Y_{n}^{\left( k \right)},\quad \forall \pi \in \mathfrak{S}_n,\label{cons_perm}
\\ 
&\mathrm{tr}_{1,\cdots,n-1}\left( Y_{n}^{\left( k \right)} \right) =Y^{\left( k \right)}, \label{eq_partialtrace}
\\ 
& Y_{n}^{\left( k \right)}\ge 0. \label{eq_refend}
    \end{align}
\end{subequations}
Here the optimization variable is $Y_n^{(k)}$ and $Y^{(k)}$ is an intermediate variable. 
$\mathfrak{S}_n$ denotes the symmetric group of degree $n$ and $U_{\pi}$ is the unitary operator that permutes the $n$ copies of subsystem $\mathbb{C}^{\mathfrak{p}+1}$ according to the permutation $\pi$. Owing to the permutation symmetry constraint (\ref{cons_perm}), the partial trace in \cref{eq_partialtrace} can be taken over any $n-1$ copies of subsystem $\mathbb{C}^{\mathfrak{p}+1}$. Here we set the partial trace to be over the first $n-1$ copies. According to the symmetric extension technique \cite{DPS02, DPS04, TPBA24}, the optimal value of \cref{eq_optrelax1} is no larger than that of \cref{eq_opt_sep} and it coincides with the latter for sufficiently large $n$. We remark that, to make the computation in Eq.~(\ref{eq_optrelax1}) more tractable, one can further impose a stronger symmetry by restricting to the symmetric subspace of $(\mathbb{C}^{\mathfrak{p}+1})^{\otimes n}$ and imposing the positive partial transpose constraints \cite{DPS04,TPBA24}.

\section{Application 1: Revisiting analytical results} \label{sec_application_1}

To illustrate the physical significance of our approach, we apply it to the three-dimensional magnetic field estimation problem. Consider a spin-$1/2$ particle subjected to a magnetic field $\boldsymbol{B}=(B_1,B_2,B_3)^T$.
The dynamics of the system is governed by the Hamiltonian $H=\sum_{i=1}^3{\mu B_i\sigma _i/2}$, where $\sigma_i$, $i=1,2,3$, denote the Pauli matrices. The coefficient $\mu$ represents the proportionality coefficient relating the spin operator $\sigma_i/2$ to the magnetic moment. From a parameter estimation perspective, the above problem is mathematically equivalent to estimating $\boldsymbol{\theta }=\left( \theta _1,\theta _2,\theta _3 \right)^T$ encoded in the Hamiltonian
$H=\theta _1\sigma _1+\theta _2\sigma _2+\theta _3\sigma _3$,
with $\theta_i = \mu B_i/2$. The associated signal channel is given by the unitary evolution 
\begin{equation}\label{eq:U}
    U_{\boldsymbol{\theta }}=e^{-iHt}
\end{equation}
where $t$ is the evolution time. We here consider the noiseless case, leaving the discussion on the noisy scenario to Sec.~\ref{sec_application_2}. The purpose of this section is to employ our approach to reveal the strengths and limitations of some analytical results, which exist for parallel and sequential strategies.

The highest achievable precision is already well understood in the noiseless case for sequential strategies. An analytical expression for this precision is derived in Ref.~\cite{Yua16} and the corresponding bound is shown to be always attainable by a physically implementable sequential strategy. In contrast, analyzing parallel strategies is much more challenging. To address this challenge, a line of research has focused on proposing some heuristic probe states and using the precision attained by these states to benchmark the highest achievable precision in parallel strategies. For example, the authors of 
Ref.~\cite{BD16} have proposed a class of permutationally invariant heuristic probe states with a two-body reduced density matrix given by 
\begin{equation}
    \rho ^{\left[ 2 \right]}=\frac{1}{4}\mathbb{I} _2\otimes \mathbb{I} _2+\frac{1}{12}\sum_{k=1}^3{\sigma _k\otimes \sigma _k}
\end{equation}
and a one-body reduced density matrix $\rho ^{\left[ 1 \right]}=\mathbb{I} _2/2$. It has been shown that the estimation error for these states is
\begin{equation} \label{eq_ana_par}
    \mathrm{tr}\left( \Sigma \right) =\frac{3}{4N\left( N+2 \right)}\left[ \frac{1}{t^2}+\frac{2\left\| \boldsymbol{\theta } \right\| ^2}{\sin ^2\left( \left\| \boldsymbol{\theta } \right\| t \right)} \right].
\end{equation}
Here $\left\| \boldsymbol{\theta } \right\|$ denotes the Euclidean norm of the vector $\boldsymbol{\theta }$, and the weight matrix $W$ is chosen to be the identity matrix. Subsequently, Ref.~\cite{HZX20} has improved upon this result by introducing a new class of heuristic probe states that achieve higher estimation precision. Nevertheless, it remains unclear whether this improved precision is the highest.

\begin{figure}[tbp]
    \centering
    \includegraphics[width=\linewidth]{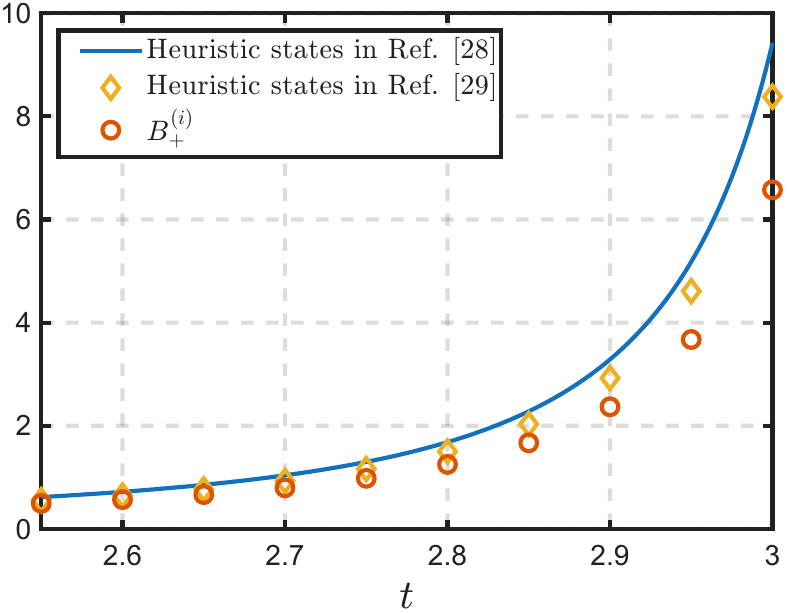}
    \caption{Estimation error as a function of $t$ for $\theta_1=\theta_2=1/2$ and $\theta_3=\sqrt{2}/2$ with $N=2$. The blue curve depicts the estimation error given by \cref{eq_ana_par}. The yellow diamonds represent the estimation error associated with the heuristic states proposed in Ref.~\cite{HZX20}. The red circles are the upper bounds on the optimal estimation error, obtained from our approach by randomly generating $m=125$ real unit vectors $\left\{ |w_x\rangle \right\} _{x=1}^{m}$.}
    \label{fig_parallel_lose_bound}
\end{figure}

With the aid of our approach, we can answer this question. As an example, we consider the parameter configuration $\theta_1=\theta_2=1/2$ and $\theta_3=\sqrt{2}/2$ and  compare the estimation error achieved by these heuristic probe states with the upper bound obtained from our approach. Hereafter, we use $B_+^{(k)}$ and $B_-^{(k)}$ to denote the upper and lower bounds obtained from our approach for the strategies of type $k$. For instance, $B_+^{(i)}$ and $B_-^{(i)}$ denote the upper and lower bounds for parallel strategies, respectively. The results are shown in \cref{fig_parallel_lose_bound}, where we plot the estimation error as a function of $t$ for $N=2$. Throughout, all numerical calculations are carried out with MOSEK \cite{MOSEK}. As can be seen from \cref{fig_parallel_lose_bound}, the estimation error achieved by all the heuristic probe states is larger than the upper bound $B_+^{(i)}$, obtained from our approach by randomly generating $m=125$ real unit vectors $\left\{ |w_x\rangle \right\} _{x=1}^{m}$. This observation indicates that these heuristic probe states are not optimal, thereby revealing the limitations of these analytical results.

\begin{figure}[t]
    \centering
    \includegraphics[width=\linewidth]{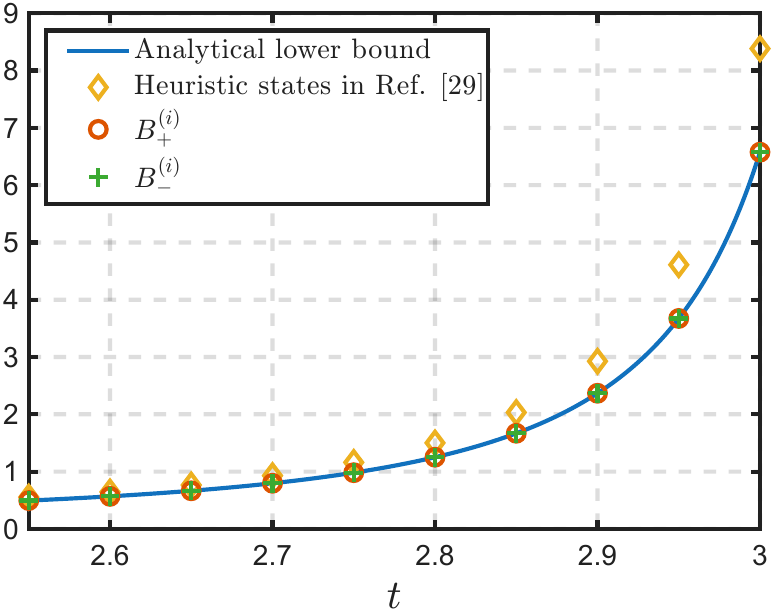}
    \caption{Illustration of the coincidence between the analytical bound in \cref{eq_ana_lower_par} and the upper/lower bounds obtained from our approach. The yellow diamonds represent the precision associated with the heuristic states proposed in Ref.~\cite{HZX20}. The blue line corresponds to the analytical bound in \cref{eq_ana_lower_par}. The red circles and green crosses denote the upper and lower bounds obtained from our approach with $m=125$ and $n=2$, respectively. Here, we set $\theta_1=\theta_2=1/2$ and $\theta_3=\sqrt{2}/2$ with $N=2$.}
    \label{fig_parallel_better}
\end{figure}

\begin{figure}[tbp]
    \centering
    \includegraphics[width=\linewidth]{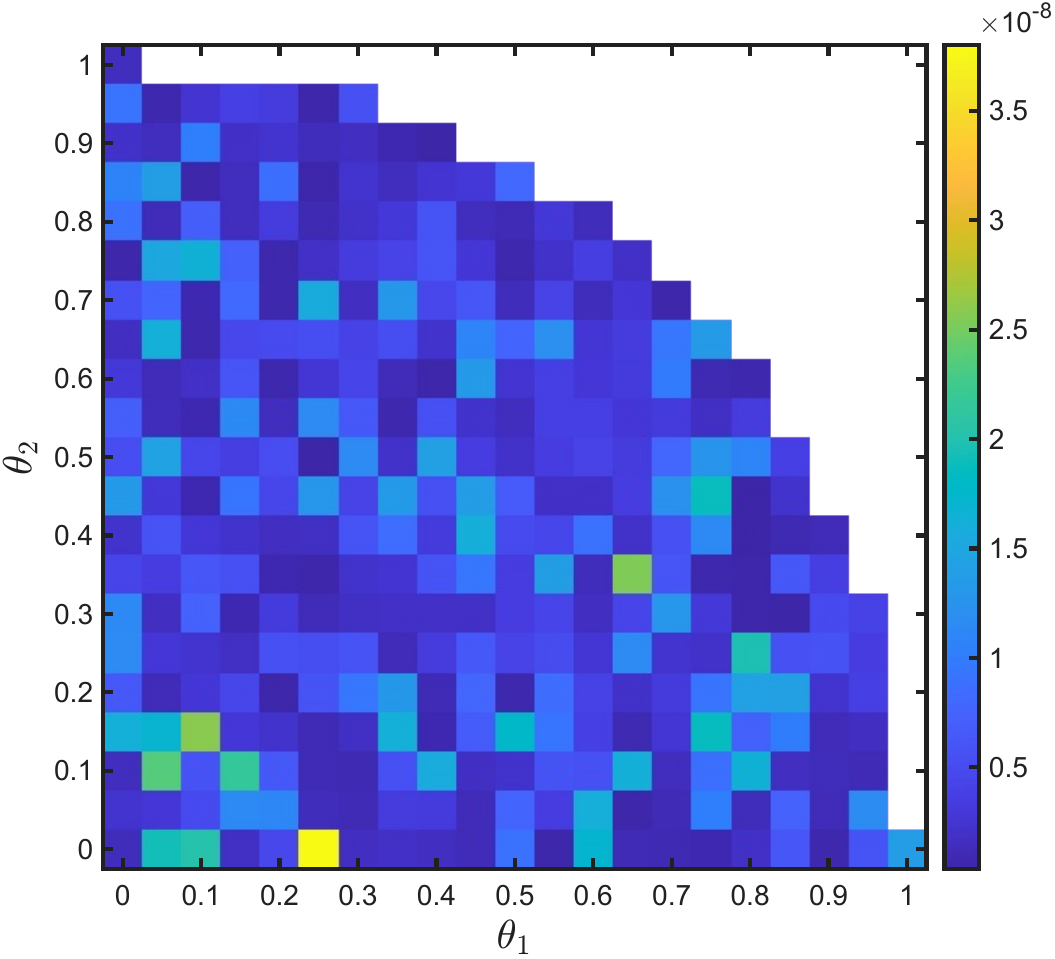}
    \caption{Differences between the analytical lower bound in \cref{eq_ana_lower_par} and the upper bound $B_+^{(i)}$ obtained from our method for different parameter configurations satisfying $\left\| \boldsymbol{\theta } \right\| =1$. Each point corresponds to a specific choice of $\theta_1$ and $\theta_2$, with $\theta_3$ determined through the relation $\theta_3 = \sqrt{1 - \theta_1^2 - \theta_2^2}$. The color denotes the magnitude of the difference. The white region represents parameter configurations that do not satisfy the constraint $\left\| \boldsymbol{\theta } \right\| =1$. Here, we set $t=3,m=700$, and $N=2$.}
    \label{fig_parallel_numerical_conjecture}
\end{figure}

We note that another notable analytical result is the analytical bound derived in Ref.~\cite{HZX20}. It states that the estimation error for parallel strategies is lower bounded by
\begin{equation} \label{eq_ana_lower_par}
    \mathrm{tr}\left( \Sigma \right) \ge \frac{\left( 1+2\frac{\left\| \boldsymbol{\theta } \right\| t}{\left| \sin \left( \left\| \boldsymbol{\theta } \right\| t \right) \right|} \right) ^2}{4N\left( N+2 \right) t^2}.
\end{equation}
This bound is asymptotically attainable in the limit of $N\rightarrow\infty$, by employing the heuristic probe states introduced in
Ref.~\cite{HZX20}. However, for small $N$, these states fail to
saturate the bound in general, as illustrated in Fig.~\ref{fig_parallel_better}. As a
result, it remains unclear whether the analytical lower bound in
\cref{eq_ana_lower_par} is tight for finite and small values of $N$. We numerically demonstrate that this bound may indeed be tight even for small $N$. 
An explicit example is shown in \cref{fig_parallel_better}, where we compare the analytical lower bound in \cref{eq_ana_lower_par} with the upper and lower bounds obtained from our approach for the parameter configuration $\theta_1=\theta_2=1/2$ and $\theta_3=\sqrt{2}/2$ with $N=2$. As depicted in \cref{fig_parallel_better}, the upper and lower bounds, $B_+^{(i)}$ and $B_-^{(i)}$, coincide with the analytical lower bound in \cref{eq_ana_lower_par}, indicating that the bound (\ref{eq_ana_lower_par}) is indeed tight for this specific case. 
To further support our observation, we have conducted extensive numerical calculations for various parameter configurations. Figure \ref{fig_parallel_numerical_conjecture} shows the differences between the analytical lower bound (\ref{eq_ana_lower_par}) and the upper bound $B_+^{(i)}$ obtained from our method for different parameter configurations satisfying $\left\| \boldsymbol{\theta } \right\| =1$. As can be seen from \cref{fig_parallel_numerical_conjecture}, these differences are nearly zero, irrespective of the specific choice of $\bm{\theta}$. We have also performed numerical calculations for other norms of $\boldsymbol{\theta}$ and observed similar behaviors. We therefore present only the results for $\left\| \boldsymbol{\theta } \right\|=1$ here. Based on these numerical results, we conjecture that the analytical lower bound in \cref{eq_ana_lower_par} remains tight even for small $N$. The rigorous proof of this statement should be an interesting topic for future work.

\section{Application 2: Establishing strict Hierarchy} \label{sec_application_2}

Let us now take noise into account. Assume that the signal channel $\mathcal{E}_{\bm{\theta}}$ is described by the unitary evolution in Eq.~(\ref{eq:U}) followed by an amplitude damping noise. This scenario is relevant in various physical platforms, such as nitrogen-vacancy centers in diamond \cite{Doherty2013PR} and superconducting qubits \cite{Kjaergaard2020ARoCMP}, where the spin-$1/2$ particles are often subject to the amplitude damping noise. Explicitly, the Kraus operators of $\mathcal{E}_{\bm{\theta}}$ are $K_1 U_{\boldsymbol{\theta }}$ and $K_2 U_{\boldsymbol{\theta }}$, with $K_1$ and $K_2$ denoting the Kraus operators for the amplitude damping noise
\begin{equation}
    K_1=\left[ \begin{matrix}
	1&		0\\
	0&		\sqrt{1-\gamma}\\
\end{matrix} \right] ,\quad K_2 = \left[ \begin{matrix}
	0&		\sqrt{\gamma}\\
	0&		0\\
\end{matrix} \right].
\end{equation}
Here $\gamma$ is the noise strength. It is worth noting that an outstanding topic in quantum metrology is to explore the hierarchy problem among different types of strategies, which has been extensively studied in the single-parameter estimation \cite{Giovannetti2006,DemkowiczDobrzanski2014,Yua16,LHYY23,KGAD23} but remains largely unexplored in the multiparameter scenario. The purpose of this section is to employ our approach to investigate this problem in the noisy multiparameter estimation scenario.

\begin{figure}[tbp]
    \centering
    \includegraphics[width=\linewidth]{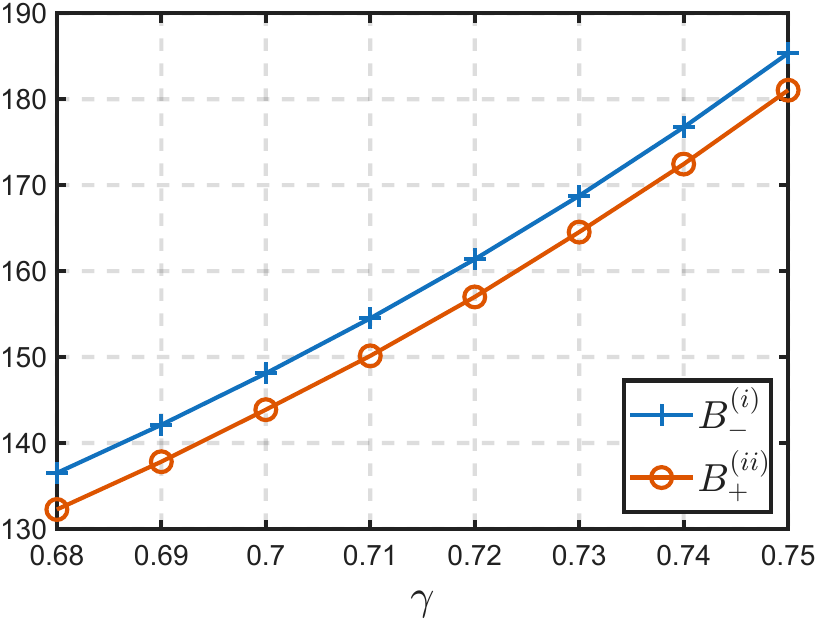}
    \caption{Lower bound $B_-^{(i)}$ for parallel strategies and upper bound $B_+^{(ii)}$ for sequential strategies as functions of the noise strength $\gamma$. Here $B_-^{(i)}$ and $B_+^{(ii)}$ are obtained from our approach by setting $n=2$ and $m=1500$, respectively. The parameters are chosen as $\theta _1=\theta _2=1/2$ and $\theta _3=\sqrt{2}/2$ with $t=0.1$ and $N=2$.}
    \label{fig_hierarchy_123}
\end{figure}

Throughout this section, we set the parameter configuration as $\theta_1=\theta_2=1/2$ and $\theta_3=\sqrt{2}/2$ with $t=0.1$ and $N=2$. With the aid of our approach, we first examine the hierarchy relation between parallel and sequential strategies. The numerical results are shown in \cref{fig_hierarchy_123}, where we plot the lower bound $B_-^{(i)}$ for parallel strategies and the upper bound $B_+^{(ii)}$ for sequential strategies as functions of $\gamma$.  Here $B_-^{(i)}$ and $B_+^{(ii)}$ are obtained from our approach by setting $n=2$ and $m=1500$, respectively. As can be seen from \cref{fig_hierarchy_123}, $B_-^{(i)}$ is strictly larger than $B_+^{(ii)}$, irrespective of the value of $\gamma$. This implies that the highest achievable precision in sequential strategies always exceeds that in parallel strategies, indicating a strict advantage of sequential strategies over parallel strategies.

\begin{figure}[tbp]
    \centering
    \includegraphics[width=\linewidth]{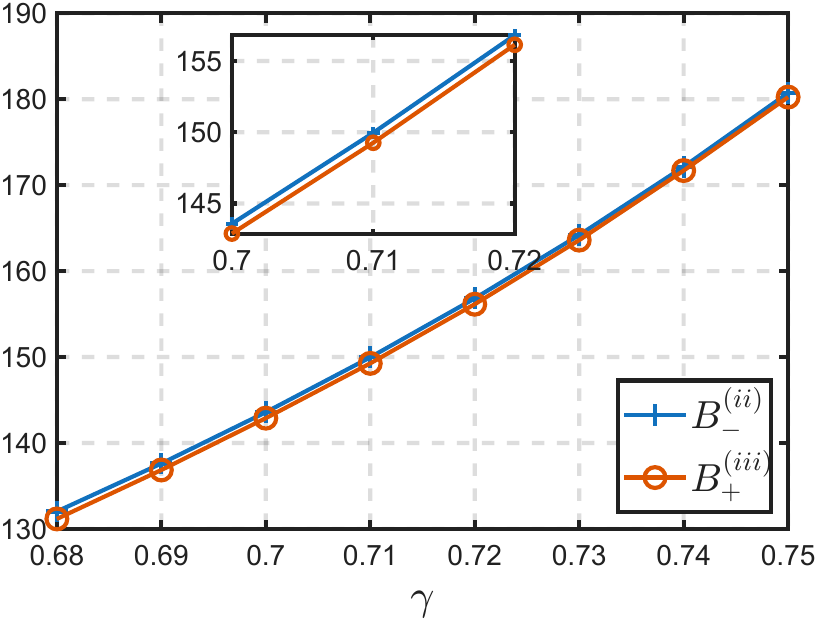}
    \caption{Lower bound $B_-^{(ii)}$ for sequential strategies and upper bound $B_+^{(iii)}$ for causal superposition strategies as functions of the noise strength $\gamma$. Here $B_-^{(ii)}$ and $B_+^{(iii)}$ are obtained from our approach by setting $n=2$ and $m=1500$, respectively. The parameters considered here are chosen to be identical to those in Fig.~\ref{fig_hierarchy_123}. The inset enlarges the gap
between the two bounds.}
    \label{fig_stri_hierarchy_23}
\end{figure}

We next examine the hierarchy relation between sequential strategies and causal superposition strategies. Figure \ref{fig_stri_hierarchy_23} shows the lower bound $B_-^{(ii)}$ for sequential strategies and the upper bound $B_+^{(iii)}$ for causal superposition strategies as functions of $\gamma$. Here $B_-^{(ii)}$ and $B_+^{(iii)}$ are obtained from our approach by setting $n=2$ and $m=1500$, respectively. Compared with the previous case (see Fig.~\ref{fig_hierarchy_123}), the gap between $B_-^{(ii)}$ and $B_+^{(iii)}$ is much smaller, as can be seen from \cref{fig_stri_hierarchy_23}. Nevertheless, there is still a strict advantage of causal superposition strategies over sequential strategies, as can be seen in the inset of \cref{fig_stri_hierarchy_23}.

\begin{figure}[tbp]
    \centering
    \includegraphics[width=\linewidth]{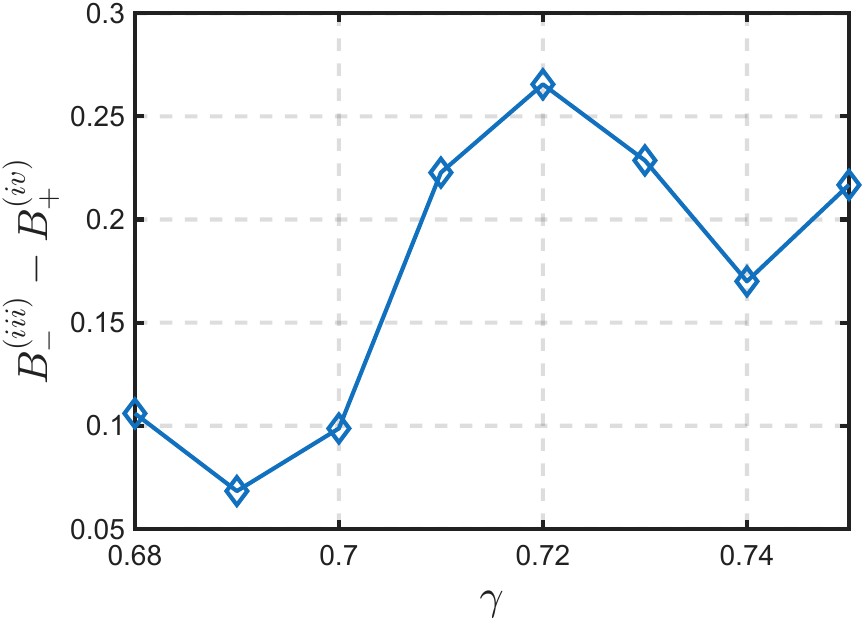}
    \caption{The difference $B_-^{(iii)}-B_+^{(iv)}$ as a function of the noise strength $\gamma$. Here $B_-^{(iii)}$ and $B_+^{(iv)}$ are obtained from our approach by setting $n=2$ and $m=1500$, respectively. The parameters considered here are chosen to be identical to those in Fig.~\ref{fig_hierarchy_123}.}
    \label{fig_hierarchy_diff}
\end{figure}

We finally investigate the hierarchy relation between causal superposition strategies and general indefinite-causal-order strategies. The numerical results are presented in \cref{fig_hierarchy_diff}, where we plot the difference $B_-^{(iii)}-B_+^{(iv)}$ as a function of $\gamma$. Here $B_-^{(iii)}$ and $B_+^{(iv)}$ are obtained from our approach by setting $n=2$ and $m=1500$, respectively. Again, we observe from \cref{fig_hierarchy_diff} that $B_-^{(iii)}$ is always larger than $B_+^{(iv)}$, indicating a strict advantage of general indefinite-causal-order strategies over causal superposition strategies. Now, combing all the results in \cref{fig_hierarchy_123,fig_stri_hierarchy_23,fig_hierarchy_diff}, we conclude that there exists a strict hierarchy among the four types of strategies in the noisy scenario.

\section{Conclusion} \label{sec_conclude}

We have presented a unified and computable approach for the simultaneous estimation of multiple parameters that attains the ultimate precision permitted by quantum mechanics. The key innovation here is to combine the quantum tester formalism and the TCRB, which together provide a unified framework for dealing with diverse quantum resources, including entanglement, coherence, quantum control, and indefinite causal order. Our approach is therefore versatile and can be applied to various types of strategies both in the noiseless and noisy scenarios.

We have clearly demonstrated the physical significance of our approach by applying it to the three-dimensional magnetic-field estimation. In the noiseless case, our numerical results identify strategies that outperform previously proposed heuristic ones and provide strong evidence that the analytical lower bound in Ref.~\cite{HZX20} is saturable even for finite and small $N$. In the noisy scenario, our approach enables a systematic comparison among different classes of strategies and establishes a strict hierarchy among parallel, sequential, causal-superposition, and general indefinite-causal-order strategies. These results extend the notion of strategy hierarchy, well understood in single-parameter estimation, to the genuinely multiparameter regime.

We clarify that our approach can accommodate arbitrary weight matrices, general noise models, and various estimation strategies. Moreover, we remark that our approach may, in principle, be extended to Bayesian estimation problems with prior distributions \cite{DGG20,MMB25}, providing a unified tool for both local and Bayesian multiparameter quantum metrology. We expect that the results presented in this work could facilitate a deeper understanding of ultimate achievable precision limits in multiparameter estimation and serve as a versatile benchmark for assessing and designing optimal quantum metrological strategies. 


\begin{acknowledgments}
    This work is supported by the National Natural Science Foundation of China (Grant No.~12275155) and the Shandong Provincial Young Scientists Fund (Grant No.~ZR2025QB16). The scientific calculations in this paper have been done on the HPC Cloud Platform of Shandong University.
\end{acknowledgments}

\section*{DATA AVAILABILITY}
The code used in this article is openly available from \cite{codelink}.

\appendix

\section{Expressions for $\Lambda^{(k)}$} \label{appsec_lambda_k}

Here we present the explicit expressions for the maps $\Lambda^{(k)}$, $k=i,ii,iv$, introduced in \cref{eq_defBBX}. In what follows, $\prescript{}{1-Q}{X^{(k)}}\coloneqq X^{(k)} - \prescript{}{Q}{X^{(k)}}$, where $\prescript{}{Q}{X^{(k)}}$ denotes the operation that traces out subsystem $Q$ and replaces it with the normalized identity operator, i.e., $\prescript{}{Q}{X^{(k)}}\coloneqq \mathrm{tr}_QX^{(k)}\otimes \left( \mathbb{I} _Q/d_Q \right)$ where $d_Q = \mathrm{dim}(Q)$. The expression for $\Lambda^{(i)}$ is 
\begin{equation}
    \Lambda^{(i)}(X^{(i)})\coloneqq\prescript{}{O_1\cdots O_N}{X^{(i)}}.
\end{equation}
The expression for $\Lambda^{(ii)}$ is
\begin{eqnarray}
    &&\Lambda^{(ii)}(X^{(ii)})\coloneqq\prescript{}{O_N}{X^{(ii)}}-
    \prescript{}{(1-O_{N-1})I_NO_N}{X^{(ii)}}-
    \\ 
    &&\prescript{}{(1-O_{N-2})I_{N-1}O_{N-1}I_NO_N}{X^{(ii)}}-\cdots-
    \prescript{}{(1-O_1)I_2O_2\cdots I_NO_N}{X^{(ii)}}.\nonumber\\
\end{eqnarray}
The expression for $\Lambda^{(iv)}$ is
\begin{equation}
    \Lambda^{(iv)}(X^{(iv)})\coloneqq\prescript{}{\left[1-\prod_{j=1}^N\left(1-O_j+I_jO_j\right)+\prod_{j=1}^NI_jO_j\right]}{X^{(iv)}}.
\end{equation}

\section{Strategies of type $iii$} \label{appsec_striii}

To simplify the notations, we focus on the case of $N=2$. All results presented here can be extended straightforwardly to arbitrary $N$. 
Denote $1\prec 2$ as the causal order such that the first channel is queried before the second channel, and similarly for $2\prec 1$. It has been shown \cite{ABC15,BMQ21,LHYY23,Zhou2024a} that the set $\mathcal X^{(iii)}$ can be characterized as 
\begin{equation} \label{appeq_cha_striii}
    \mathcal{X} ^{\left( iii \right)}\coloneqq \left\{ X|X=pX^{\left( 1\prec 2 \right)}+\left( 1-p \right) X^{\left( 2\prec 1 \right)},0\le p\le 1 \right\},
\end{equation}
where $X^{\left( 1\prec 2 \right)}\in \mathcal X^{(ii)}$ has causal order $1\prec 2$ and $X^{\left( 2\prec 1 \right)}\in \mathcal X^{(ii)}$ has causal order $2\prec 1$. 
With the characterization of $\mathcal{X}^{(iii)}$ given in \cref{appeq_cha_striii}, the upper bounds for causal superposition strategies can be written explicitly as
\begin{subequations} \label{appeq_opt_upper_iii}
    \begin{align}
       \underset{\begin{array}{c}
	 X_{x}^{\left( iii \right)} \\
	\tilde X^{\left( 1\prec 2 \right)},\tilde X^{\left( 2\prec 1 \right)}\\
\end{array}}{\min}\,\,&\sum_x{\langle w_x|\tilde{W}|w_x\rangle \mathrm{tr}\left( C_{\boldsymbol{\theta }}X_{x}^{\left( iii \right)} \right)}
\\
\mathrm{s}.\mathrm{t}.\,\,&\sum_x{\left| \langle w_x|0\rangle \right|^2X_{x}^{\left( iii \right)}}=\tilde X^{\left( 1\prec 2 \right)}+\tilde X^{\left( 2\prec 1 \right)},
\\ 
& \Lambda ^{\left( 1\prec 2 \right)}\left( \tilde X^{\left( 1\prec 2 \right)} \right) =\tilde X^{\left( 1\prec 2 \right)}, \label{appeq_X12X21_cons1}
\\ 
& \Lambda ^{\left( 2\prec 1 \right)}\left( \tilde X^{\left( 2\prec 1 \right)} \right) =\tilde X^{\left( 2\prec 1 \right)}, \label{appeq_X12X21_cons2}
\\
& \mathrm{tr}\left[ \tilde X^{\left( 1\prec 2 \right)}+\tilde X^{\left( 2\prec 1 \right)} \right] =d_O, \label{appeq_X12X21_cons3}
\\ 
& \tilde X^{\left( 1\prec 2 \right)}\ge 0,\quad \tilde X^{\left( 2\prec 1 \right)}\ge 0, \label{appeq_X12X21_cons4}
\\ 
&X_{x}^{\left( iii \right)}\ge 0,\quad \forall x,
\\
&\sum_x{\langle w_x|A_i |w_x\rangle \mathrm{tr}\left( \partial _jC_{\boldsymbol{\theta }}X_{x}^{\left( iii \right)} \right)}=\delta _{ij},\nonumber
\\
&\quad \quad \quad \quad \quad \quad \quad \quad \quad \quad \quad \quad i,j=1,\cdots ,\mathfrak{p},
    \end{align}
\end{subequations} 
where the variable $p$ in \cref{appeq_cha_striii} has been absorbed into $\tilde X^{\left( 1\prec 2 \right)}$ and $\tilde X^{\left( 2\prec 1 \right)}$, i.e., $\tilde X^{\left( 1\prec 2 \right)} = p X^{\left( 1\prec 2 \right)}$ and $\tilde X^{\left( 2\prec 1 \right)} = (1-p) X^{\left( 2\prec 1 \right)}$. Inserting \cref{appeq_cha_striii} into \cref{eq_optrelax1}, we can express the lower bounds for the causal superposition strategies as 
\begin{subequations}
    \begin{align}
        \underset{\begin{array}{c}
	Y^{(iii)}_n\\
	\tilde X^{\left( 1\prec 2 \right)},\tilde X^{\left( 2\prec 1 \right)}\\
\end{array}}{\min}\,\,& \mathrm{tr}\left[ \tilde{W}\otimes C_{\boldsymbol{\theta }}Y^{(iii)} \right] 
 \\ 
 \mathrm{s}.\mathrm{t}.\,\, &\mathrm{tr}_{1}\left[ |0\rangle \langle 0|\otimes \mathbb{I} _{IO}Y^{(iii)} \right] =\tilde X^{\left( 1\prec 2 \right)}+\tilde X^{\left( 2\prec 1 \right)},
 \\ 
 &\text{Eqs.~(\ref{eq_refbegin}-\ref{eq_refend}) hold},
 \\ 
 & \text{Eqs.~(\ref{appeq_X12X21_cons1}-\ref{appeq_X12X21_cons4}) hold.}
    \end{align}
\end{subequations}


\begin{thebibliography}{59}
\bibitem [{Not()}]{Note}%
  \BibitemOpen
  \href@noop {} {}\bibinfo {note} {The same authors have studied how to find the optimal input state for parallel strategies in Ref.~\cite{HO24}, which can be considered as a special case addressed in this work.}\BibitemShut {Stop}%
\end{thebibliography}

%
\end{document}